# A PMU-Based Machine Learning Application for Fast Detection of Forced Oscillations from Wind Farms

Mohammed-Ilies Ayachi [1], Luigi Vanfretti [2] and Shehab Ahmed [1]

[1] Physical Science and Engineering Division, King Abdullah University of Science and Technology, Saudia Arabia
[2] Department of Electrical, Computer and Systems Engineering, Rensselaer Polytechnic Institute, USA

*Abstract--* Today's evolving power system contains an increasing amount of power electronic interfaced energy sources and loads that require a paradigm shift in utility operations. Sub-synchronous oscillations at frequencies around 13-15 Hz, for instance, have been reported by utilities due to wind farm controller interactions with the grid. Dynamics at such frequencies are unobservable by most SCADA tools due to low sampling frequencies and lack of synchronization. Real-time or off-line frequency domain analysis of phasor measurement unit (PMU) data has become a valuable method to identify such phenomena, at the expense of costly power system data and communication infrastructure. This article proposes an alternative machine learning (ML) based application for sub-synchronous oscillation detection in wind farm applications. The application is targeted for real-time implementation at the 'edge', resulting in significant savings in terms of data and communication requirements. Validation is performed using data from a North American wind farm operator.

## I. INTRODUCTION

*A) Motivation*

Renewable energy's rapid growth and penetration in today's power system is creating new challenges: rapid voltage fluctuations, transient / sub-synchronous oscillations, damping degradation, and other operational issues. Low inertia power systems receiving a mixture of instantaneous non-synchronous renewables such as wind turbines and solar photovoltaics, are subject to transient stability issues that have already been studied [1]. While oscillations are always present in power systems and typically do not represent a major threat to power system stability, they can often indicate incipient instabilities or serious equipment problems. In some cases, forced oscillations can result in resonant interactions and equipment damage [2]. Forced oscillations arise from equipment malfunction or controller interactions, and can take on many forms including sinusoids, limit cycles, or just erratic signals; an example of recorded PMU measurements is shown in Fig. 1.

Real-time monitoring of oscillations is critical to the reliable and safe power system operation. Devices called phasor measurement units (PMU) measure voltage and current phasors, and with these measurements, PMUs derive frequency estimates. These PMU measurements provide real-time grid data to operators, which helps them make decisions to prevent outages.

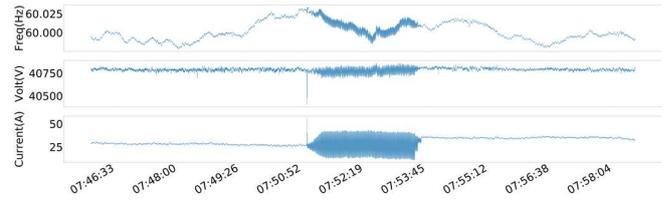

*Fig. 1. Example of oscillation at 1.7Hz, duration 00:02:37, recorded by PMU, resulting from interactions between wind farms*

Recently, transmission system operators have used PMUs to measure sub-synchronous oscillatory events resulting from interactions between wind farms at frequencies around 13-15 Hz [3]. Oscillations reached the consumer level in the form of flicker [4]. For many cities, new energy plans are leading to an increase in renewable and distributed energy resource penetration, rendering the detection and possible prevention of similar oscillations as critical. Ignoring such events is not an option due to the uncertainties of the changing energy landscape.

*B) Previous work*

Oscillation detectors have been around for 30 years. However, a system trigger for disturbance monitoring based on lowpass filters tends to occasionally miss an important disturbance that is just within the range of the monitor since the software will only issue an alarm based on the oscillation trigger if a threshold is attained for a predefined amount of time [5]. In [6] and [7], this algorithm is expanded to use PMU data and four different RMS energy filters. In these approaches, the software triggers an alarm if active power changes are large enough. Furthermore, the oscillation must also persist for a pre-determined time for the application to issue an alarm. Also, in [8], a fast real-time oscillation detection preprocesses the data with outlier removal and down sampling using different filters (bandpass, highpass, and low pass) with trigger flags after level comparison. A LabVIEW based software allows us to set up the different threshold values and frequencies to survey.

Together with detection, the classification of the type of disturbance that gives rise to oscillations is also of interest. In [9], a MATLAB-based software has been developed for disturbance classification and identification using a decision tree based on 30 seconds of dynamic recorded data and a hardcoded threshold. In [10] and [11], four types of disturbances are identified by processing PMU frequency data through deep neural networks, using real data from Brazil, with a time window of 20 seconds. In [12] supervised machine-learning for event detection using FFT transforms PMU data and reduces its input dimensionality through the Principal Component Analysis process.

*C) Contributions of this paper*

The main contributions of this paper are:

(i) To propose a new oscillation detection method that exploits raw PMU data without any need for specialized signal processing, such as noise removal and filtering, and without the need for hardcoded thresholds.
(ii) To propose a software implementation that requires only 1 sec of data for real-time application and that can be deployed at the edge of the network on a small Linux-based IoT device.
(iii) To provide a comparison between different machine learning algorithms for purposes of oscillation detection.
(iv) To propose the use of transfer-learning by generating the training data using a Modelica-based simulation model and determining the impact of injecting real data during the inference phase.

## II. OSCILLATION DETECTION AND EVENT CLASSIFICATION

Data from PMUs are transported via a communication network using the IEEE C37.118.2 protocol. In [13] a proposed implementation to extract voltage, current, frequency, Rate-Of-Change-Of-Frequency, and other user-defined analog and digital state data is given. Fig. 1 shows the raw data from a PMU during a timeline of 10 minutes with a clearly visible oscillation in frequency, voltage, and current. Using 1 second of this data obtained during and outside the oscillation event, the frequency is plotted as shown in Fig. 2.

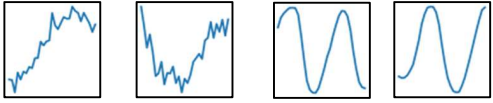

*Fig. 2. Four examples of frequency plotting (one second each), during normal condition and during oscillation, from real PMU data.*

Plotting voltage or current gives similar results. With a sampling rate of 30 samples per second, this particular oscillation is easy to distinguish.

Machine learning in power systems has more than 20 years of history, it is being used in several sub-areas of power engineering, for example, load forecasting, but it is still not being used for power system operations in the production environment. The reasons for this lack of adoption are complex and include a lack of understanding on how to apply ML algorithms, the availability and pipelining of data, vendor lock-in of software technology and its' lack of fitness for deployment in modern computer hardware architectures for ML.

With the rise of ML-capable IoT devices, like the Nvidia's Jetson product line, image classification in real-time at the edge is becoming possible. The only difficulty remains in obtaining access to real data records to train and verify the ML model during the inference phase. For the problem addressed herein, sub-synchronous wind farm oscillation detection, modeling, and simulation software (Modelica/Dymola) allows us to generate precisely the data needed to train the ML model.

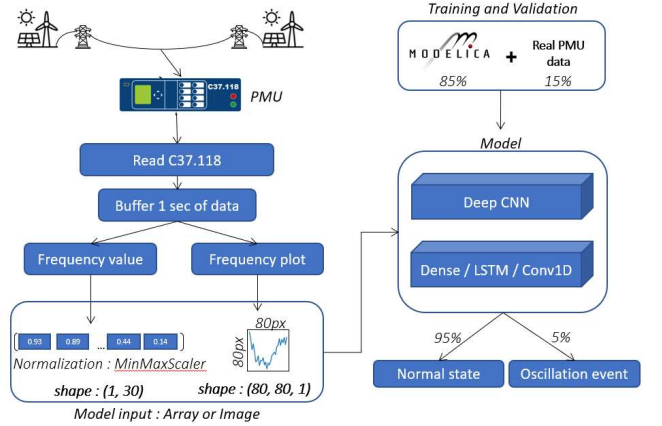

*Fig. 3. Workflow applied to the data: an array of raw frequency data is input to the Dense/LSTM model; an image of the frequency plotting for CNN model*

## III. METHODOLOGY

In this section, we describe the methodology for oscillation detection using ML in two parts. Firstly, we describe the generation of training data and thereafter present the approach for applying ML algorithms. Our goal, as shown in Fig. 3, is to train a model with synthetic data, and then continue the training with a small proportion of real data. Depending on the type of model, the input will be an array of normalized frequencies or a picture of the frequency plot.

*A) Generating Training Data*

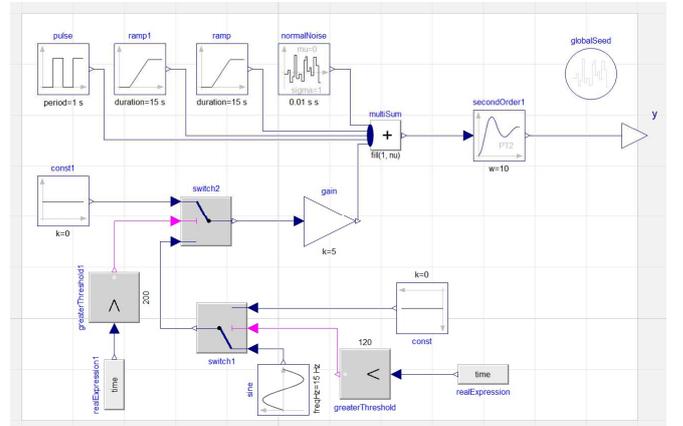

*Fig. 4. Modelica model diagram used in Dymola to generate the synthetic training data*

In Fig. 4 a simulation model to generate synthetic data like the one in Fig. 1 is presented. Two ramps and a pulse signal are used to simulate a progressive increase and decrease of the amplitude of the signal, while a noise generator with a normal distribution is used to model the stochastic behavior. A sine signal with a defined frequency, activated by two switches in a specified time delta, is added to the input of a second-order transfer function which is used to represent low-frequency dynamics. The result from plotting one second of the generated signal is shown in Fig. 5 which shows that this approach allows the generation of seamlessly similar time-series of a sub-synchronous wind farm oscillation observed on real data coming from a PMU.

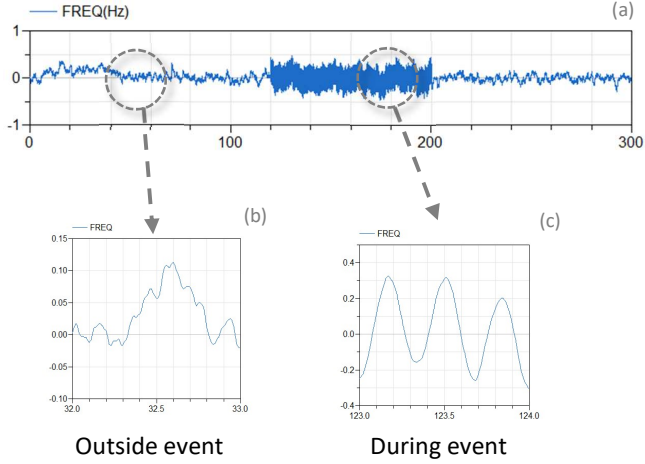

*Fig. 5. On top (a): 300 seconds of Dymola output, 3 Hz oscillation from 120 to 200 sec. On bottom : 1 second zoom outside (b) and during oscillation event (c).*

To be able to deploy the oscillation detector on an IoT device (like a Raspberry Pi, Beaglebone Black or Nvidia Jetson) at the edge of the power system infrastructure (Fog computing), the software needs to be executed on a Linux environment. The Python language was used for the whole development due to its convenience, built-in data structures and practicality. To manipulate the Modelica/Dymola model's variable with Python and directly get the desired outputs, the model was first exported to FMU (Functional Mock-up Unit) and imported with the python library FMPy. The sinusoidal frequency is changed from 1Hz to 15Hz for each export, and the output is plotted on a time delta of 1 second. Exports of the events during oscillation are merged in the same folder so that ML training is done with only two folders of pictures (inside and outside event), as shown in Fig. 6.

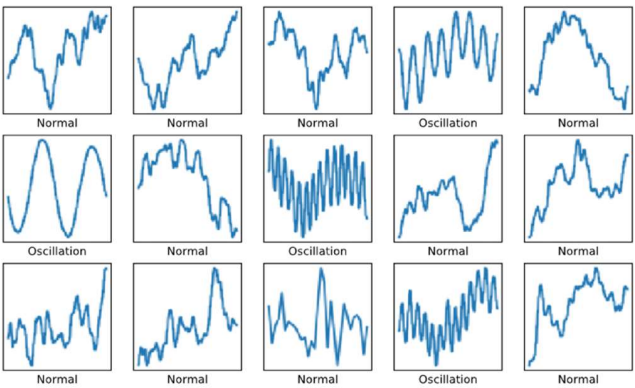

*Fig. 6. Matrix of the two classes of data generated during and outside of an event. Oscillation sinusoid varies from 1Hz to 15Hz*

To apply machine learning methods to both simulation and real measurement data, we chose to use the Tensorflow and Keras frameworks. Among different methods, Convolutional Neural Networks (CNN) are classifiers that are known to have outstanding performances in the field of pattern recognition with an image input. In this work, the input is an image of the frequency plotting, with a size of 80x80 pixel. A similar approach was used for ECG arrhythmia classification, using data transformation and augmentation to feed a deep CNN, see [14] and [15] . Fig. 7 is an illustration of how CNN patterns are learned through translation: after learning a shape, a convnet can recognize it anywhere. The first convolution layer learns small local patterns; a second convolution layer will learn larger patterns made of the features of the first layers, and so on.

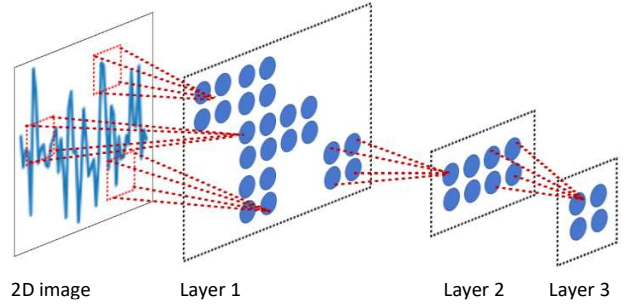

*Fig. 7. Illustration of the action of a CNN over an image: segmentation and progressive merger.*

The CNN model architecture implemented for this work closely mirrors AlexNet's architecture [16] with less depth. As shown in Table 1, ReLU is the activation function used to bring non-linearity, and the final layer uses a sigmoid activation to output a probability between 0 and 1. Because oscillation detection is a binary classification problem, and the output of the network is a probability, a categorical cross-entropy loss is used [17] .

| Layer | No. of channels | Activation |
|---|---|---|
| Convolution 2D | 64 | ReLU |
| Convolution 2D | 64 | ReLU |
| Batch Normalization | | |
| Convolution 2D | 64 | ReLU |
| Max Pooling 2D | (2,2) | |
| Convolution 2D | 64 | ReLU |
| Convolution 2D | 64 | ReLU |
| Batch Normalization | | |
| Max Pooling 2D | (2,2) | |
| Convolution 2D | 32 | ReLU |
| Convolution 2D | 32 | ReLU |
| Convolution 2D | 32 | ReLU |
| Max Pooling 2D | (2,2) | |
| Dropout | 0.3 | |
| Flatted | | |
| Dense | 512 | ReLU |
| Dense | 256 | ReLU |
| Dropout | 0.5 | |
| Dense | 2 | Softmax |

*Table 1. Layers and parameters of the Convolutional Neural Network designed*

We performed a comparison between the performance of the proposed CNN and that of other well-known architectures: MobileNet, ResNet and AlexNet. Thanks to the generated training data, an unlimited amount of data is available, and therefore, there is no need for data augmentation. To illustrate how an input decomposes into the different filters learned by the network, we have shown a feature map of our second convolution layer in Fig. 8.

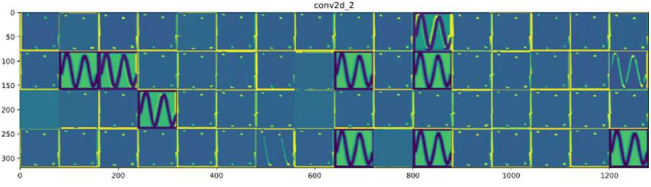

*Fig. 8. Visualization of the intermediate activations displaying the feature maps that are produced by the second convolution of the proposed CNN*

Our proposed CNN utilizes 2D convolutions, extracting 2D patches from image tensors and applying an identical transformation to every patch. It is possible, in the same way, to use 1D convolutions to extract local 1D patches (subsequences) from sequences. 1D convnets can be competitive with RNNs on sequence-processing problems at a considerably cheaper computational cost. Recently, 1D convnets have been used with great success for audio generation and machine translation [17] . In addition to these specific successes, it has long been known that small 1D convnets can offer a fast alternative to RNNs for simple tasks such as timeseries forecasting. Our Conv1D model is composed of two Conv1D of 64 channels, one MaxPooling1D followed by a Dropout, then one Dense layer of 100 channels (Table 2). Detection accuracy is almost identical to the CNN, but computational resources are divided by two.

| Layer | No. of channels | Activation |
|---|---|---|
| Convolution 1D | 64 | ReLU |
| Convolution 1D | 64 | ReLU |
| Max Pooling 1D | (2) | |
| Dropout | 0.5 | |
| Flatten | | |
| Dense | 100 | ReLU |
| Dropout | 0.5 | |
| Dense | 2 | Softmax |

*Table 2. Our proposed 1D convnet*

## IV. RESULTS

To evaluate the performance of the proposed oscillation detector, we have used recordings of real data coming from PMUs from a wind farm in North America. From the entire dataset of events, several records have clearly visible oscillations with a sufficiently long time duration, which allows us to distinguish it from a false positive event, as shown in Fig. 9 .

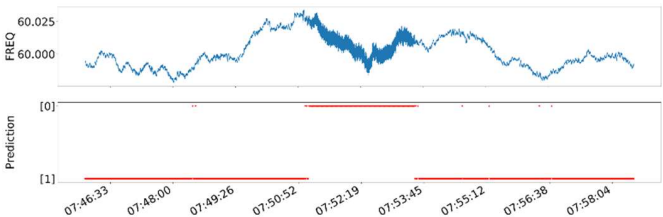

*Fig. 9. Proposed CNN Model (trained using simulated and real data) predicting each second if the frequency inputted (coming from a real PMU recording) is an oscillation event or not : 1 if normal event and 0 if oscillation event.*

The PMU's data used can contain up to 446 different terminals, each one geographically placed on a different transmission line, leading to 2 Gb of data per csv file. Some of the transmission lines do not reflect the oscillations at all. To assess accuracy, we manually selected and used for training one transmission line that exposes clear oscillations. The results are shown in Table 3. Accuracy is the number of good predictions (any class) divided by the total number of samples (size of dataset). To avoid having a biased result by having more normal events than oscillation events, we truncated the dataset used so as to contain exactly 2 minutes of normal events and 2 minutes of oscillation events (Fig. 10).

| Model | Accuracy | False-positive | Missed event | Time for 1 prediction (sec) |
|---|---|---|---|---|
| Proposed CNN | 97.41% | 2 | 6 | 0.0047 |
| Proposed Conv1D | 98.06% | 0 | 6 | 0.0027 |
| MobileNet | 97.74% | 2 | 5 | 0.0074 |
| MobileNet[2] | 98.71% | 0 | 4 | 0.0074 |
| AlexNet | 94.51% | 12 | 5 | 0.0098 |
| ResNet-50 | 97.42% | 4 | 4 | 0.0174 |
| Dense | 94.19% | 6 | 12 | 0.0026 |
| Stacked LSTM | 94.19% | 2 | 16 | 0.0054 |

*Table 3. Accuracy and performance comparison between our CNN / Conv1D and other well-known architecture. ( (2) = with ImageNet weights loaded before inference)*

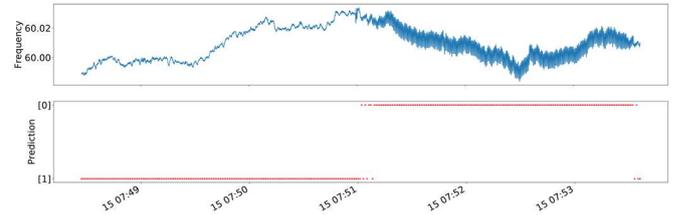

*Fig. 10. Conv1D predicting on 4 minutes of datas: 2 min of normal followd by 2 min of oscillation. Sliding window of 1 second.*

To test the Dense, stacked LSTM and Conv1D architecture, instead of using an image as an input, an array of 30 normalized raw frequency values (real data have a sample rate of 30 samples per second), are passed to the model. The Dense architecture is composed of a Dense layer with 256 and 64 units with a dropout in the middle. The LSTM model is a stack with the shape of 512-256-128-64-32 units. Both models have a SoftMax activation and a sparse categorical loss function.

The proposed CNN architecture shows a global accuracy of 97.41% and the proposed Conv1D shows an accuracy of 98.06% . The time to make a prediction is the average based on 1000 predictions made on different hardware. During development, a Windows PC with a Core i7-8700 CPU and Nvidia 1080ti graphics card with Tensorflow 2.0-GPU was used, while two other IoT devices were used to compare edge device performance as summarized below (Table 4).

| Hardware | Time for 1 prediction with CNN | Time for 1 prediction with Conv1D |
|---|---|---|
| Windows PC Core i7 8700 – Nvidia 1080Ti | 0.0049 sec | 0.0022 sec |
| Nvidia Jetson Xavier | 0.0357 sec | 0.0170 sec |
| Raspberry Pi 3 | 0.4698 sec | 0.0114 sec |

*Table 4. Average mean inference time based on 1000 samples*

Even on a small device without GPU computational capabilities like the Raspberry PI 3, the model can predict an output under a second, allowing its use as a real-time monitoring application on cost-effective devices deployed at the edge.

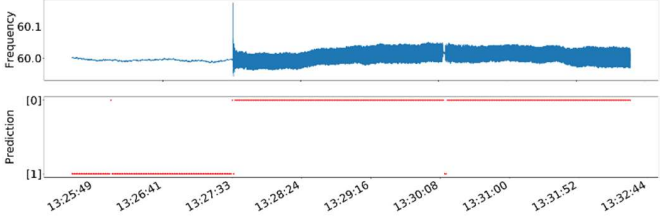

*Fig.11. Our CNN Model trained on Dymola's data only predicting with a sliding window of 1 seconds.*

Depending on the selected transmission line, the impact of adding real data during the training phase of the model (transfer learning) may lead to substantial prediction improvements. For example, in Fig. 11, the model trained only on synthetic data performs remarkably well. On another dataset of real PMU data, several misclassifications during the oscillation are present, as shown in Fig. 12 .

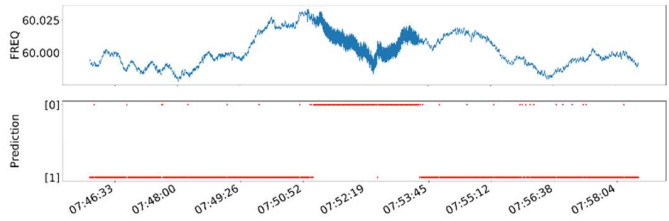

*Fig. 12. Model trained on synthetic data only predicting with a sliding window of 1 second: there is a lot of false positive event.*

This implies that it is necessary, for certain measurement locations, to train the model with real data. Training the model with real data is made with 3 selected terminals, from 3 different CSV files corresponding to 3 different days where oscillation is visible. To be able to isolate the impact of injecting real data during the model training phase, 73 terminals with clearly visible oscillations were selected. However, we observe that some of the false-positive event detected by the CNN may be actual oscillation events. A comparison of the accuracy during the oscillation event between a model trained on synthetic data only and a model trained using both synthetic and real data is shown in Fig. 13.

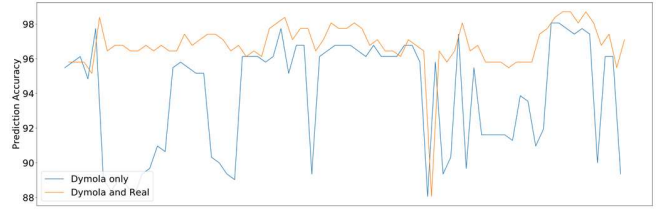

*Fig. 13. Accuracy of the oscillation detection method with a model trained on synthetic data only (blue) and a model trained on real + synthetic data (orange). Each point is a different terminal (transmission line).*

The transfer learning approach shows better results than using synthetic data only. The average prediction accuracy measured on 73 terminals for the model trained with simulation data only is 93.94%, and for the model trained on simulation + real data, it is 96.79%. The simulation dataset consists of 11964 files for the training and 2991 files for validation (25% ratio). The real data contains 967 samples (equivalent to 15% of the synthetic dataset).

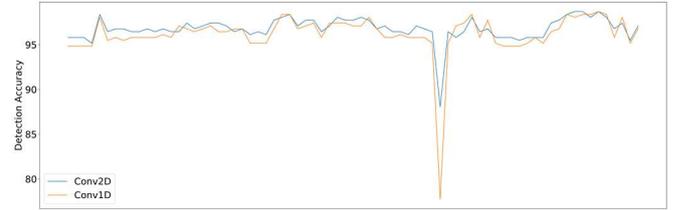

*Fig. 14. Global accuracy measured on 73 terminals for CNN (Conv2D) vs Conv1D*

Experimentation shows a very small advantage for 2D CNN, most of the time with a global accuracy of 96.79% for Conv2D, over 96.19% for Conv1D (2 minutes of normal events then 2 minutes of oscillation events) as shown in Fig. 14.

## V. SUMMARY AND FUTURE WORK

We have presented an approach for fast detection of forced oscillation using one second of buffered PMU data with a deep CNN and a 1D convnet machine learning approach. The software can be deployed on low cost Linux based devices, and only utilizes open-source publicly available libraries. Experimental results indicate that we were able to achieve high accuracy oscillation event detection running in real-time on cost-effective hardware. Our current model classifies its input into two classes (normal condition and oscillation event). An additional useful feature would be to estimate the exact frequency of the oscillation. Training the model to distinguish each oscillation frequency, from 1Hz to 15Hz, and have 16 outputs instead of only 2 has resulted in preliminary positive results. Identifying the exact oscillation frequency with high precision using a regression estimator will be studied in future work. We modeled this oscillation detection as a classification task, future work could also model it as a "temporal detection" task in order to avoid the few misclassified points by learning from the surrounding context information, that is similar to approaches using in video understanding research [18].

## Biographies

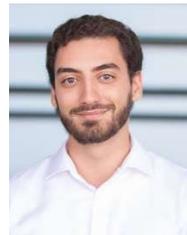

**Mohammed-iliès Ayachi**, M.S. software engineering from University of Strasbourg, France. He worked 6 years as a software engineer in France before joining Kaust as a research specialist in October 2018 on the development of new tools inside the eco-system of Smart grid, in both the energy sector and sustainable development.

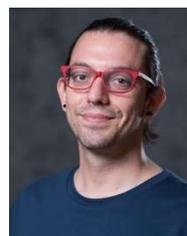

**Luigi Vanfretti** obtained the MSc and PhD degrees from Rensselaer Polytechnic Institute in 2007 and 2009. He was with KTH Royal Institute of Technology from 2010-2017 and with Statnett SF from 2013-2016. Previous research includes the development of a software platform that uses machine learning, currently used by RTE, France.

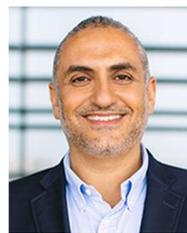

**Shehab Ahmed**, received the B.Sc. degree in electrical engineering from Alexandria University in 1999, and the M.Sc. and Ph.D. degrees from the Department of Electrical and Computer Engineering, Texas A&M University, College Station, TX in 2000 and 2007, respectively. He was with Schlumberger Technology Corporation, Houston, TX from 2001 to 2007 developing downhole mechatronic systems for oilfield service products. He was with Texas A&M University from 2007 to 2018. He is currently Professor of Electrical Engineering with the CEMSE Division at King Abdullah University of Science and Technology.